\documentclass[%
 aip,
 jmp,%
 amsmath,amssymb,
 reprint,%
]{revtex4-1}

\usepackage{graphicx}
\usepackage{dcolumn}
\usepackage{bm}
\usepackage{sidecap}
\usepackage{ifthen}

\begin{document}

\preprint{AIP/123-QED}

\title{Quasiresonant Excitation of InP/InGaP Quantum Dots Using Second Harmonic Generated in a Photonic Crystal Cavity}

\author{Sonia Buckley}
\email{bucklesm@stanford.edu.}
 \affiliation{E. L. Ginzton Laboratory, Stanford University, Stanford, CA 94305, U.S.A.}

\author{Kelley Rivoire}
\affiliation{E. L. Ginzton Laboratory, Stanford University, Stanford, CA 94305, U.S.A.}

\author{Fariba Hatami}
\affiliation{Department of Physics, Humboldt University, D-12489, Berlin, Germany}

\author{Jelena Vu\v{c}kovi\'{c}}
\affiliation{E. L. Ginzton Laboratory, Stanford University, Stanford, CA 94305, U.S.A.}


\begin{abstract}
Indistinguishable single photons are necessary for quantum information processing applications.  Resonant or quasiresonant excitation of single quantum dots provides greater single photon indistinguishability than incoherent pumping, but is also more challenging experimentally.  Here, we demonstrate high signal to noise quasiresonant excitation of InP/InGaP quantum dots.  The excitation is provided via second harmonic generated from a telecommunications wavelength laser resonant with the fundamental mode of a photonic crystal cavity, fabricated at twice the quantum dot transition wavelength.  The second harmonic is generated using the $\chi^{(2)}$ nonlinearity of the InGaP material matrix.
\end{abstract}

\pacs{42.70.Qs, 78.67.Hc, 42.65.Ky}
\maketitle

\begin{figure*}[t]
\includegraphics[width=18cm]{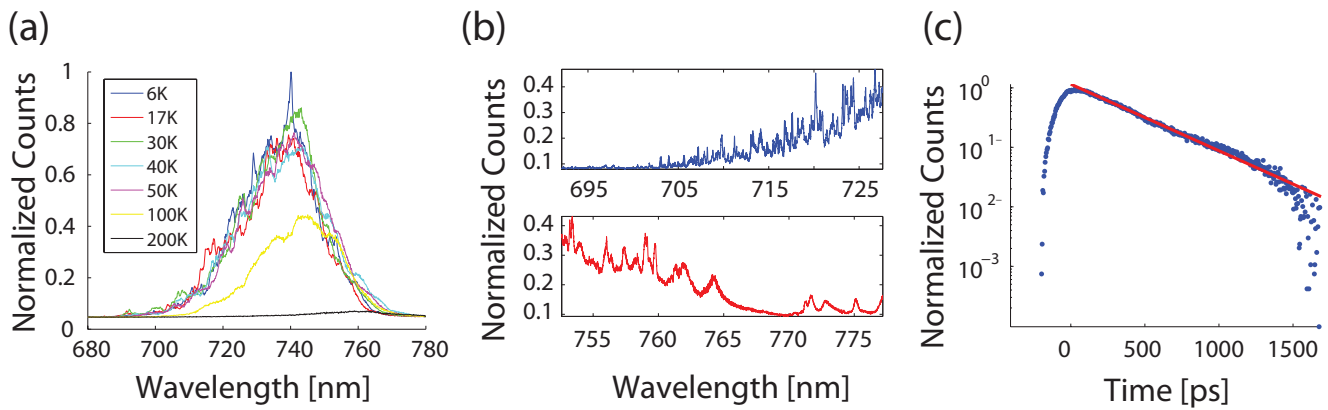}
\caption{Characterization of InP/InGaP QD properties.  (a) Temperature dependence of QD photoluminescence under excitation with a CW 405 nm diode laser at 70 $\mathrm{\mu}$W. (b) QD distribution at 6 K taken with 1714 groove/mm grating with 20 $\mathrm{\mu}$W excitation power, showing comparison of QD spectrum at the short (blue) and long (red) wavelength sides of the distribution (c) Time-resolved streak-camera measurements showing the lifetime integrated across the PL spectrum. The red line shows an exponential decay $\tau$ = 390 ps. \label{fig:DotProperties}}
\end{figure*}

Single semiconductor quantum dots (QDs) at visible wavelengths are beneficial for implementing single photon sources for quantum applications, since Si avalanche photodiodes (APDs) have maximum detection efficiencies in the red part of the spectrum \cite{eisaman_invited_2011}. For most of these applications (including quantum computing), photon indistinguishability is critical  for maintaining the high fidelities required.  Additionally, emission in the visible part of the spectrum can be frequency downconverted to telecommunications wavelengths using readily available lasers \cite{zaske_efficient_2011}, in contrast to InAs/GaAs QDs which generally emit at wavelengths of around 900 nm (and thus frequency conversion to telecommunications wavelength in this case requires a pump laser at around 2.3 $\mathrm{\mu}$m).  Clear single quantum emitters in the red spectral range with narrow emission lines exhibiting antibunching have been observed only in the InP/InGaP\cite{ugur_single-dot_2008, ugur_single-photon_2012} and InP/AlGaInP QD systems \cite{schulz_optical_2009}, and from non-QD systems such as nitrogen vacancy centers in diamond \cite{beveratos_room_2002}. Site-controlled InP/InGaP QDs have also been fabricated recently \cite{baumann_site-controlled_2012}, which are important for scalable applications.

Above-band excitation provides the simplest way of exciting single QDs, as the excitation and fluorescence will be at different frequencies and can be spectrally filtered efficiently. However, in this case the single photon indistinguishability is lost due to dephasing from the  solid state environment and timing jitter resulting from relaxation of carriers and their capture in QDs \cite{santori_indistinguishable_2002}.  Resonant and quasi-resonant excitation help to improve this indistinguishability, as the carrier capture process is eliminated, and phonon assisted carrier relaxation is minimized.  Quasi-resonant and resonant excitation have been demonstrated in the InP/InGaP system previously \cite{vollmer_exciton_1996,eberl_preparation_1995} for QD samples without resolvable single QD lines, using a Ti:Sapphire laser in a cross polarized configuration for excitation and using time resolution to filter the remaining scattered laser from the signal. Here we demonstrate a novel method for achieving quasi-resonant excitation of a QD/photonic crystal cavity system using second harmonic light generated in the photonic crystal cavity to excite the QDs.  The second harmonic is generated using the $\chi^{(2)}$ nonlinearity of the surrounding InGaP material, and is enhanced by the resonant photonic crystal cavity mode.  This technique has a high signal to noise ratio and has the advantage of using a compact and low power telecommunications laser as a pump. Scattered light from the high power pump laser is spectrally separated from the emitted QD photoluminescence (PL), while unwanted nonlinear processes, such as Raman scattering of the pump into the quasi-resonantly excited dot distribution, will not affect the signal.  The resonant excitation is provided by intracavity second harmonic generation (SHG), which results in significantly less scattered light, with outcoupled intensity of the same order of magnitude as the PL from the QD.  Other schemes for resonant excitation involve strong polarization filtering \cite{vamivakas_spin-resolved_2009} or using a cavity to suppress scattered light \cite{flagg_resonantly_2009}.  Nonlinear excitation allows flexible excitation wavelength over a broad wavelength range, and in principle this could be extended by using sum frequency generation in cavities with multiple resonances \cite{rivoire_sum-frequency_2010}.

\begin{figure}[h!]
\includegraphics[width=9cm]{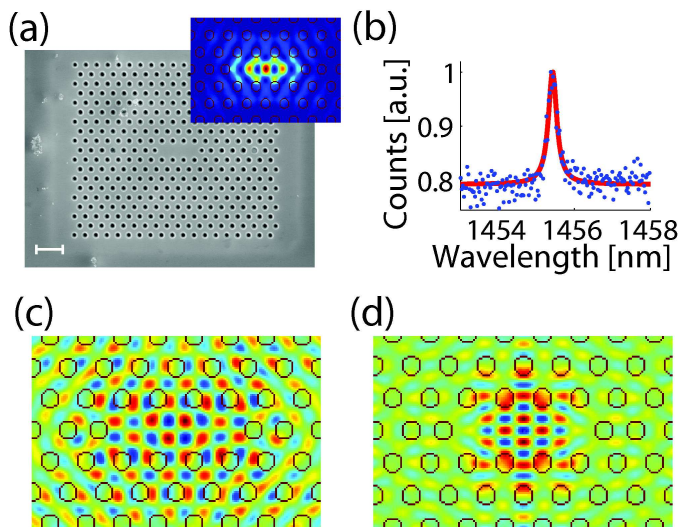}
{\caption{Characterization of photonic crystal cavity. (a) SEM of the photonic crystal cavity. Inset shows $|E|$ calculated by FDTD simulation. (b) Cross polarized reflectivity spectrum of the fundamental resonance of the photonic crystal cavity. Red line is a Lorentzian fit showing a Q of 7000. (c) $E_z$ of the TM mode of the second harmonic generated in the cavity at twice the fundamental cavity frequency. (d) $E_x$ of a TE mode to which the QD PL couples. This mode is at roughly 4\% longer wavelength than the TM mode. \label{fig:PC}}}
\end{figure}

The QDs are grown by gas-source molecular-beam epitaxy (GSMBE) in the center of a 150-nm-thick In$_{0.48}$Ga$_{0.52}$P membrane grown on top of a 500-nm layer of Al$_{0.8}$Ga$_{0.2}$As on a (001) GaAs substrate. The areal density of the QDs is 3-5 per $\mathrm{\mu}$m$^2$. We first characterized the QD properties before fabricating our photonic crystal samples. Fig. \ref{fig:DotProperties} (a) shows the measured QD PL as a function of temperature from 6 K to 200 K in a continuous-flow helium cryostat using 70 $\mathrm{\mu}$W excitation power from a 405-nm continuous-wave (CW) diode laser (this power was chosen as convenient for measuring at all temperatures). The center wavelength of the QD emission redshifts by 20 nm from 6 K to 200 K. The gaussian shape and large full width half maximum of the emission results from inhomogeneous broadening due to variation in the physical size of the QDs \cite{wu_effect_1987}. Fig. \ref{fig:DotProperties} (b) shows the QD PL for two different spectral windows in the QD distribution with the same laser diode with 20 $\mathrm{\mu}$W excitation power, taken at 6 K with a finer grating (1714 grooves/mm) to resolve the single QD lines.  Single lines can be seen from approximately 700-750 nm; QDs at the blue side of the distribution exhibit narrower line widths and are more efficiently excited by SHG than the QDs at the red side.  However, the high density of these spectral lines makes quantitative measurements of the linewidth difficult. We investigate the dynamics of the ensemble-QD emission by studying the time-resolved photoluminescence on a streak camera when the QDs are excited at 400 nm by a frequency-doubled Ti:Sapphire laser with a repetition rate of 80 MHz and integrating over the higher energy half of the PL spectrum of the dots. The data was taken for the shorter wavelength half of the spectrum, as the best quasi-resonant excitation was demonstrated from dots on this side of the distribution.  The time resolved data is shown in Fig. 1 (d)  plotted on a semi-log scale and fitted with an exponential (red line) with time constant $\tau$ = 390 ps, while a longer wavelengths yielded shorter time constants, with 300 ps measured for the long wavelength part of the spectrum.  The different properties of QDs in different parts of the emission spectrum were reported previously \cite{christ_carrier_2000}, however, in that study the lifetime of the QDs was shown to increase with wavelength.  The falloff from an exponential distribution, seen at long times, may be due to carrier relaxation from the above band excitation \cite{eberl_preparation_1995}. The short time constant is consistent with previous measurements of this QD system \cite{eberl_preparation_1995, christ_carrier_2000, vollmer_exciton_1996}.\\

\begin{SCfigure*}
\includegraphics[width=12.5cm]{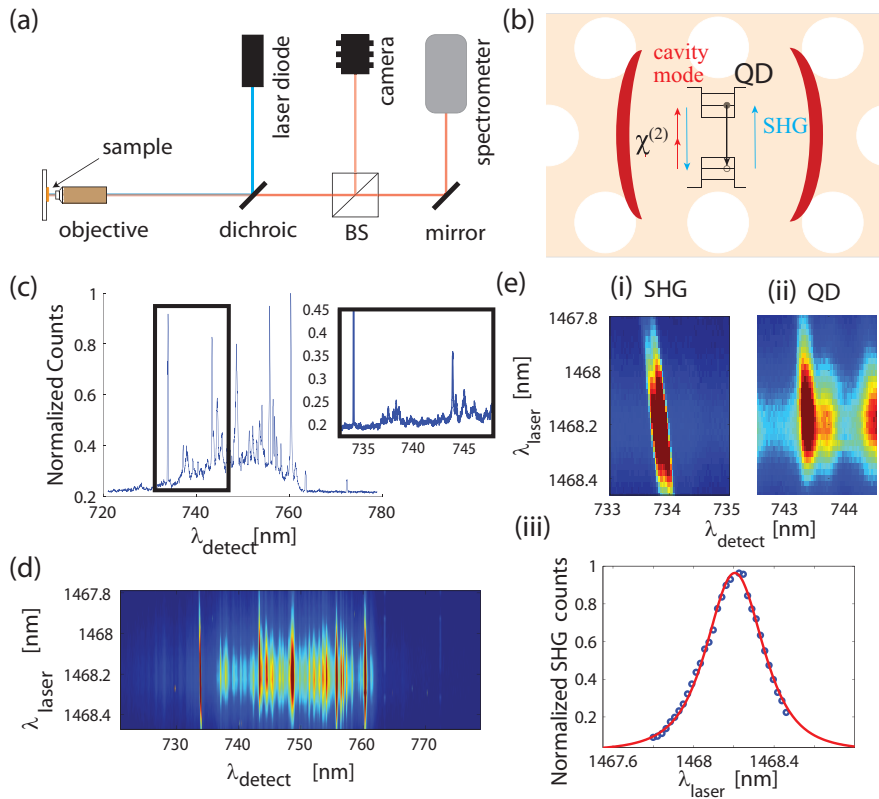}
\caption{(a) The setup used in the experiment. BS = beamsplitter. (b) Schematic of intracavity second harmonic generation and excitation of a QD. (c) A single spectrum with pump wavelength at 1468.1 nm taken with 600 g/mm grating. Inset shows spectrum taken with 1714 g/mm grating. (d) The emission intensity as a function of pump and detected wavelengths.  The cavity second harmonic is the diagonal line around 734 nm. (e) Zoom in on (i) second harmonic and (ii) a QD line from part (d).  Part (iii) shows intensity of second harmonic versus pump laser wavelength, with a fit to a Lorentzian squared.  No evidence of resonant excitation is seen.
\label{fig:QuasiResonant}}
\end{SCfigure*}

 Fig. \ref{fig:PC} (a) shows a scanning electron microscope (SEM) image of the three-hole linear defect photonic crystal cavity, fabricated at half the QD frequency ($\sim$ 1500 nm wavelength), such that the second harmonic will resonantly excite the QD. The structures are fabricated in the InGaP membrane using e-beam lithography, dry etching, and HF acid wet etching of the sacrificial layer beneath the membrane, as described previously \cite{rivoire_gallium_2008}. The cavity main axis is oriented along a 100 crystal direction.  Finite difference time domain (FDTD) simulations of the electric field components were performed and the magnitude of the electric field is shown in the inset of Fig. \ref{fig:PC} (a). The resonance is TE-like with electric field primarily in the plane of the photonic crystal slab.  By varying hole size and other fabrication parameters, photonic crystal cavities with resonances from 1452 nm to 1569 nm were fabricated.  These were characterized using a broadband source in the cross-polarized configuration, as in our previous work \cite{altug_polarization_2005}, to maximize signal to noise.  A reflectivity spectrum of a cavity measured by this technique and showing a resonance at 1455.5 nm is shown in Fig. \ref{fig:PC} (b). The red line shows a Lorentzian fit of this spectrum, and gives a cavity Q of 7000. Light coupled into this resonance can be frequency doubled by second harmonic generation \cite{mccutcheon_experimental_2007,rivoire_second_2009} using the $\chi^{(2)}$ nonlinearity in the non-centrosymmetric InGaP material. Due to the symmetry of the photonic crystal, the resonance has dominant E$_x$ and E$_y$ in-plane field components. The $\chi^{(2)}$ tensor of the InGaP has symmetry such that the only non-zero component is $\chi^{(2)}_{i\neq j\neq k} \neq 0$.  This means that the TE-like mode with fields primarily in the $x$ and $y$ polarizations must accordingly generate second harmonic that is primarily z-polarized, i.e., a transverse magnetic-like (TM-like) mode \cite{rivoire_second_2009}. For a photonic crystal, this corresponds to guided resonances in the TM-like air-band. The TM mode at the frequency of the second harmonic was simulated using FDTD, and this component of the electric field in the out of plane ($z$) direction, $E_z$ is shown in Fig. \ref{fig:PC} (c).  The mode was excited in simulation with a narrow frequency excitation with the polarization $P_z$ generated from the fundamental L3 mode multiplied by the effective $\chi^{(2)}$ tensor. The generated second harmonic couples into this mode at twice the fundamental frequency (around 700-750 nm), and is used to quasi-resonantly excite the QDs. The Q of this mode is 170, with 37\% of the light emitted into the light cone, and around 70\% of this is radiated into the NA of the objective lens (0.75 NA).  Anisotropy of the QD causes the dot to emit primarily TE polarization \cite{schmidbauer_shape-mediated_2002}; at the QD wavelength (roughly 10nm shorter than the SH excitation wavelength) the photonic crystal supports an optical airband TE-like mode.  The electric field in the $x$ direction for such a mode is shown in Fig. \ref{fig:PC} (d); for this mode $E_y$ has odd symmetry and so outcouples poorly. This mode was simulated using a TE excitation at the center of the slab.  The Q of this mode is 295, with 60\% of the energy emitted into the light cone (about 70\% of this is emitted into the NA of our objective lens, 0.75). This suggests a slight preference for the TE mode to outcouple from the slab over the TM mode, which adds to the signal to noise ratio of this method, although, as discussed later, the QD may be excited more efficiently by the TE polarization, especially in the case of resonant excitation.  \\

In order to demonstrate quasiresonant excitation, the setup shown in Fig. \ref{fig:QuasiResonant} was used. A tunable continuous wave (CW) telecommunications wavelength laser Agilent 81989A, 1463-1577 nm is coupled into the fundamental high-Q mode of the photonic crystal cavities. Light from the laser is reflected from a dichroic and coupled to the cavity at normal incidence with an objective lens. Photons upconverted in the cavity through second harmonic generation quasi-resonantly excite the QD, as shown in Fig. 3 (b).  The QD emission is collected through the same objective and passes through the dichroic to be measured on a liquid nitrogen cooled Si CCD spectrometer.  A QD spectrum obtained by this method from a cavity with resonant wavelength at 1468 nm can be seen in Fig. \ref{fig:QuasiResonant} (c), taken with a 600 groove/mm grating.  The shortest wavelength peak is the second harmonic.  It was verified that both the second harmonic and QD PL follow a quadratic dependence on the laser power, as in previous work \cite{rivoire_second_2009}. The inset black box shows a spectrum taken with a 1714 groove/mm grating.   A 2D plot of the intensity of the PL against the pump and detected wavelengths is shown in Fig. \ref{fig:QuasiResonant} (d). The second harmonic is the diagonal line from 733.9 to 734.2 nm detected wavelength while the other vertical lines indicate quasi-resonantly excited QDs. Fig. \ref{fig:QuasiResonant} (e) part (i) shows a zoom-in of the SHG, while part (ii) shows a zoom in of a QD line.  As the laser is scanned across the cavity, the intensity of SHG follows a Lorentzian squared, shown in Fig. (e) part (iii); the red line is a Lorentzian squared fit to the data.  The laser wavelength was tuned in steps of 0.002 nm.  The lifetime limited linewidth of these QDs is 0.005 nm (around 2 GHz), and in practice experimentally measured resonantly excited linewidths have shown broadening relative to this limit \cite{flagg_resonantly_2009}; therefore we expect that a resonantly excited QD would be observable by this method. The linewidth remains a Lorentzian squared at all scanned wavelengths and powers, and demonstrates no broadening from the laser linewidth, which is on the order of MHz, indicating that either no resonant excitation occurs or that it is much weaker than the out-coupled second harmonic.   That no evidence of resonantly excited QDs was found could be due to the strong polarization of this QD system \cite{schmidbauer_shape-mediated_2002} and thus low overlap of the resonant transition with the TM excitation. The density of QDs in this experiment was too high (3-5 per $\mathrm{\mu}$m$^2$) to spectrally isolate single QD lines.  Repeating these experiments using a lower density QD sample would allow further characterization of the quasiresonant process, for example by allowing quantitative comparisons between quasiresonant and above-band pumping, as well as measurement of the linewidth of single QDs.\\

In conclusion, we have demonstrated high signal to noise quasi-resonant excitation of InP/InGaP QDs coupled to photonic crystal cavities.  The excitation was provided via second harmonic generated from a telecommunications wavelength laser in a photonic crystal cavity using the $\chi^{(2)}$ nonlinearity of the InGaP material matrix. The method is high signal to noise as the second harmonic that excites the QD is generated within the slab, minimizing scattered light at the wavelength of the QD.  Using sum frequency generation, it could be extended to give greater flexibility of the excitation wavelength.  A QD system with less anisotropy could show resonant excitation.  The technique could prove useful as a convenient method of providing indistinguishable photons from a solid state source, which is necessary for quantum information processing.\\
\\
This work was supported by the National Science Foundation (NSF Grant ECCS-
10 25811), a National Science Graduate Fellowship, and Stanford Graduate Fellowships.  This work was performed in part at the Stanford Nanofabrication Facility of NNIN supported by the National Science Foundation under Grant No. ECS-9731293, and at the Stanford Nano Center.

\end{document}